# Giant optical birefringence of semiconductor nanowire metamaterials

*Otto L. Muskens[1*], Maarten H. M. van Weert[1], Magnus T. Borgström[2], Erik P.A.M. Bakkers[2], Jaime Gómez Rivas[1]*

[1] FOM Institute for Atomic and Molecular Physics AMOLF, c/o Philips Research Laboratories, High Tech Campus 4, 5656 AE, Eindhoven, The Netherlands.

[2] Philips Research Laboratories, High Tech Campus 4, 5656 AE, Eindhoven, The Netherlands

muskens@amolf.nl

**Semiconductor nanowires exhibit large polarization anisotropy for the absorption and emission of light[1], making them ideal building blocks for novel photonic metamaterials. Here, we demonstrate that a high density of aligned nanowires exhibits giant optical birefringence, a collective phenomenon observable uniquely for collections of wires. The nanowire material was grown on gallium phosphide (GaP) (111) in the form of vertically standing GaP nanowires. We obtain the largest optical birefringence to date, with a difference between the in-plane and out-of-plane refractive indices $\Delta n$ of 0.80 and a relative birefringence $\Delta n/n_\perp$ of 43%. These values exceed by a factor of 75 the natural birefringence of quartz and a by more than a factor of two the highest values reported so far in other artificial materials.[2] By exploiting the specific crystallographic growth directions of the nanowires on the substrate, we further demonstrate full control over the orientation of the optical birefringence effect in the metamaterial.**

For individual nanowires, large anisotropy of the absorption and emission of light has been demonstrated by polarization-dependent luminescence measurements on a single InP nanowire.[1] The anisotropy of the single wires was explained from the large difference in the polarizability of a thin cylinder for polarizations of the light parallel and perpendicular to the cylinder axis. For high densities of aligned nanowires, the anisotropy of the individual wires may result in large differences in the propagation speed of light along the different directions and polarizations of the material, a phenomenon that has not yet been reported in nanowires. Birefringence resulting from an arrangement of anisotropic objects, known as form-birefringence, has been observed in various systems such as aligned carbon nanotube films[3], porous silicon[2,4], and polymer mirrors.[5] Its magnitude depends critically on several parameters, such as the refractive index contrast, the volume fraction, alignment, and the topology (direct versus inverted medium) of the mixture. The highest birefringence values to date were measured in the mid-infrared for an anisotropically-etched silicon 2D photonic crystal with an optimized pore ratio.[2] The obtained refractive index difference of 0.366 was limited by the inverted structure of the photonic crystal. Since semiconductor nanowires generate much higher depolarization fields than air pores in a semiconductor host material, a much larger birefringence of up to 0.9 is predicted for the direct topology of the nanowire material.[6]

In order to obtain large optical anisotropy, we use the semiconductor with the highest refractive index in the visible, gallium phosphide (GaP), for the nanowires to achieve the largest possible index contrast.[7] GaP has an energy bandgap of 2.26 eV ($\lambda_{gap}$= 548 nm) and a refractive index around 3.3 for light of wavelength longer than $\lambda_{gap}$.[8] Vertically oriented nanowires were grown in a high volume density of several percent in the vapor-liquid-solid (VLS) growth mode[9] by the use of metal-organic vapor phase epitaxy (MOVPE) at an elevated temperature of 420°C. Trimethylgallium and phosphine were used as precursors using hydrogen as carrier gas. A schematic picture of the subsequent growth steps is shown in Fig. 1(a), for a detailed description of the growth parameters we refer to Ref. 10. We used a (111)-oriented GaP substrate with both sides mechanically and chemically polished to optical quality. On the P-terminated (111)B side, a thin gold film was deposited with a thickness of approximately 0.2 nm. Such a thin layer of gold breaks up into small islands resulting in individual catalyst droplets at the deposition temperature. After the nanowires were grown and to enhance the volume fraction of semiconductor material, GaP layer was radially grown on the wires at higher temperature (630°C) where the kinetic hindrance for growth on the sidewalls is overcome.

Figure 1(b) shows a scanning electron micrograph of a nanowire metamaterial fabricated using a combination of VLS and radial growth. The metamaterial layer consists of a very high density of well-aligned nanowires with a micrometer length. In Fig. 1(c-e), cross-sectional SEM images are shown from three representative nanowire samples with increasing radial growth times, (a) no shell growth, (b) 500 s growth time, and (c) 950 s growth time. It can be seen that longer radial growth times lead to thicker nanowires with roughly equal lengths of around 1.4 µm. The radial growth of GaP on the sidewalls of the nanowires results in an increase of the average nanowire diameter from 21.5 nm for the material of Fig. 1(a), 45 nm for Fig. 1(b), and up to 90 nm for Fig. 1(c). From these diameters we estimate a concomitant increase in semiconductor volume fraction by more than an order of magnitude over the range of samples under study. The increase in wire thickness is accompanied by a clear transition from a transparent to an opaque yellowish appearance of the composite medium due to the increase of light scattering[7]. The average volume density of nanowires, denoted as $f$, was determined from gravimetrical measurements and top-view SEM images to $f < 0.04$ for the material without shell up to $f = 0.40 \pm 0.05$ for the sample with the longest radial growth time.

Optical birefringence describes the difference in refractive index for different polarization directions of light traveling inside an anisotropic material. For the material consisting of aligned nanowires, the directions perpendicular to the wire axis (in-plane polarization) have a lower refractive index than the direction parallel to the nanowires (out-of-plane polarization).[11] A good description in the long-wavelength limit and up to high volume densities is given by the Maxwell Garnett effective medium approximation, taking into account the effect of the surrounding dipoles on the polarizability of a single wire[6]. The alignment of the nanowires perpendicular to the surface leads to a strong variation of the optical birefringence between normal and in-plane incidence of light. To determine the material refractive indices, we therefore use the method of angular-dependent polarization interferometry at an optical wavelength of 632.8 nm (He:Ne laser). The polarization of the input light beam is set to 45° with respect to the plane of incidence. The rotation of the polarization vector after propagation through the nanowire layer is measured via the parallel ($I_{//}$) and cross-polarized ($I_\perp$) intensities (relative to the input polarization) using an analyzing polarization filter. The ratio of the intensities $I_\perp / I_{//}$ is a measure for the achieved polarization extinction and represents the quality of the birefringent material.[4] This ratio is shown in Fig. 2 as a function of incident angle $\theta_{in}$ for light reflected from three samples with increasing nanowire volume fraction. For the presented nanowire materials, a sequence of maxima and minima is observed, corresponding to an optical retardation over multiple orders of $\pi$. Since the intrinsic birefringence of the GaP crystal itself is negligible[12], the observed effect can be unambiguously assigned to form-birefringence of the nanowire material. Accurate estimates of the in-plane and out-of-plane refractive indices were obtained by fitting the polarization-dependent reflectivities using a transfer matrix model accounting for the various interfaces and for the birefringence of the nanowire layer (lines in Fig. 2). To substantiate these results, we have performed additional angle-dependent transmission measurements as well as spectrally resolved ellipsometry at a fixed angle.[13] The values for the in-plane and out-of-plane refractive indices obtained by fitting all these independent measurements agree well to within several percent. The resulting birefringence values $\Delta n$ for all the nanowire materials with increasing radial growth times are plotted versus the semiconductor volume fraction $f$ in Fig. 3 (open dots). The birefringence increases strongly with the volume fraction. We obtain values as high as $0.80 \pm 0.07$, which is the highest reported birefringence to date. The closed diamond at a volume fraction of 0.45 indicates the highest value obtained for an inverted structure consisting of a two-dimensional lattice of air pores in silicon.[2] It agrees with the predictions from the Maxwell Garnett model for a porous network medium (dashed line in Fig. 3). For the nanowire samples (black line in Fig. 3), the Maxwell Garnett model predicts the enhanced birefringence as observed in our experiments. A parameter of considerable interest is the birefringence normalized to the in-plane refractive index $n_\perp$, describing the relative difference of the phase velocity of the light between the two polarizations. This relative birefringence reaches up to a surprisingly high value of 43% (see inset Fig. 3). The large relative birefringence is specific for the nanowire metamaterial, due to the very strong suppression of the nanowire polarizability perpendicular to its axis[1], and very different from porous network materials (red line in inset Fig. 3).

Finally, we demonstrate control over the direction of the anisotropy using the specific crystallographic growth directions of the GaP nanowires. The nanowires preferentially grow along the (111)B direction. Therefore, by growing from a GaP (100) substrate it is possible to obtain nanowire ensembles with twofold symmetry in the plane. Figure 4(a) shows cross-sectional SEM images of GaP nanowires grown from a GaP (100) substrate, for a section cleaved along the (110)-plane. We observe a high-density distribution of nanowires with a large fraction of the wires under an angle of around 35º. Conclusive evidence of this preferential nanowire orientation is given by the optical birefringence experiment presented in Fig. 4(b,c). Figure 4(c) shows the polarization contrast of the transmitted light for light incident normal to the sample, for rotation of the sample over its azimuthal axis $\phi$. The fourfold symmetry is a result of the in-plane anisotropy of the nanowire layer, analogous to the symmetry of a conventional waveplate retarder with an in-plane optical axis. The strong modulation of the contrast from $10^{-3}$ to a maximum of 0.1 demonstrates the alignment of a sizeable fraction of the nanowires in the material.

In conclusion, we have reported giant optical birefringence from a photonic metamaterial based on semiconductor nanowires, and have demonstrated full control over its properties using radial growth and the specific orientation of nanowires on the surface. By growing very-high density nanowire metamaterials that show by far the largest optical birefringence to date, we have made a significant step forward to large-scale application of nanowires. Although we focused our investigation on gallium phosphide nanowires, these novel nanowire materials can be grown using any group III-V, II-VI, or IV semiconductor on any crystalline surface. A great advantage of the bottom-up fabrication of the photonic nanowire metamaterials is the high degree of control over local growth[14] and their compatibility with silicon technology[15,16], which is directly applicable to important applications in micro- and nanophotonics.

Acknowledgements

We thank George Immink, Eddy Evens, Frans Holthuysen, and Peter Breijmer for technical assistance and Gert 't Hooft, Jeroen Kalkman, and Ad Lagendijk for discussions. This work was supported by the Netherlands Foundation "Fundamenteel Onderzoek der Materie (FOM)" and the "Nederlandse Organisatie voor Wetenschappelijk Onderzoek (NWO)", and is part of an industrial partnership program between Philips and FOM.

[1] J. Wang, M. K. Gudiksen, X. Duan, Y. Cui, C. M. Lieber, Highly polarized photoluminescence and photodetection from single indium phosphide nanowires, *Science* 293, 1455 (2001)

[2] F. Genereux, S. W. Leonard, H. M. Van Driel, A. Birner, U. Gösele, Large birefringence in two-dimensional silicon photonic crystals, *Phys. Rev. B* 63, 161101:1-4 (2001)

[3] W. A. de Heer, W. S. Bacsa, A. Chatelain, T. Gerfin, R. Humphrey-Baker, L. Forro, D. Ugarte, Aligned nanotube films: production and optical and electronic properties, *Science* 268, 845-846 (1995)

[4] N. Künzner, D. Kovalev, J. Diener, E. Gross, V. Y. Timoshenko, G. Polisski, F. Koch, M. Fujii, Giant birefringence in anisotropically nanostructured silicon, *Opt. Lett.* 26(16), 1265-1267 (2001)

[5] M. F. Weber, C. A. Stover, L. R. Gilbert, T. J. Nevitt, A. J. Oudekirk, Giant birefringence optics in multilayer polymer mirrors, *Science* 287, 2451-2456 (2000)

[6] A. Kirchner, K. Busch, C. M. Soukoulis, Transport properties of random arrays of dielectric cylinders, *Phys. Rev. B* 57, 277-288 (1998)

[7] F. J. P. Schuurmans, D. Vanmaekelbergh, J. van de Lagemaat, A. Lagendijk, Strongly photonic macroporous gallium phosphide networks, *Science* 284, 141-143 (1999)

[8] E. D. Palik, Handbook of optical constants of solids, 1st ed.; Academic Press: Orlando, 445-449 (1985)

[9] R. S. Wagner, W. C. Ellis, Vapor-liquid-solid mechanism of single crystal growth, *Appl. Phys. Lett.* 4, 89-90 (1964)

[10] M. A. Verheijen, G. Immink, T. de Smet, M. T. Borgström, E. P. A. M. Bakkers, Growth kinetics of heterostructured GaP-GaAs nanowires, *JACS* 128, 1353-1359 (2006)

[11]     M. Born, E. Wolf, *Principles of Optics*; 6$^{th}$ ed.; Cambridge University Press: Cambridge, 665-680 (1997)

[12]     J. H. Burnett, Z. H. Levine, E. L. Shirley, Intrinsic birefringence in calcium fluoride and barium fluoride, *Phys. Rev. B* 64, 241102:1-4 (2001)

[13]     O. L. Muskens, M. H. M. van Weert, M. T. Borgström, E. P. A. M. Bakkers, J. Gómez Rivas, to be published

[14]     A. I. Hochbaum, R. Fan, R. He, P.Yang, Controlled growth of Si nanowire arrays for device integration, *Nano Lett.* 5, 457-460 (2005)

[15]     E. P. A. M. Bakkers, J. A. Van Dam, S. De Franceschi, L. P. Kouwenhoven, M. Kaiser, M. Verheijen, H. Wondergem, P. Van der Sluis, Epitaxial growth of InP nanowires, *Nature Materials* 3, 769-773 (2004)

[16]     T. Mårtensson, C .P. T. Svensson, B. A. Wacaser, M. W. Larsson, W. Seifert, K. Deppert, A. Gustafsson, L. R. Wallenberg, L. Samuelson, Epitaxial III-V nanowires on silicon, *Nano Lett.* 4, 1987-1990 (2004)

**Figure 1**. (a) Scheme of semiconductor nanowire growth using the vapor-liquid solid (VLS) method starting from small gold catalyst droplets on the surface of a GaP (111)B substrate, followed by controlled radial growth of material on the walls of the nanowires. (b) SEM image of GaP nanowire material grown using a combination of VLS and radial growth. Scale bar denotes 5 μm. (c-e) Cross-sectional SEM images of GaP nanowire metamaterials with radial growth times of (c) 0 s, (d) 500 s, and (e) 950 s. Scale bars denote 1 μm.

**Figure 2.** Reflection of an impinging light beam on the slab of aligned nanowire layers using incident light polarization at 45º and detection of cross-polarized ($I_\perp$) and parallel-polarized ($I_{//}$) components. Resulting reflectivity extinction ratios $I_\perp/I_{//}$ are shown against angle of incidence $\theta_{in}$ for three nanowire materials with average nanowire diameters of 22 nm (black), 34 nm (red), and 45 nm (blue). Experimental points (Open symbols) and model fits (lines) using transfer matrix model with uniaxial nanowire birefringence.

**Figure 3.** Values $\Delta n$, obtained from fitting of the angular-dependent measurements, against nanowire volume fraction $f$ for samples with different radial growth times. Black line denotes Maxwell Garnett effective medium approximation for a pillar medium using a refractive index for the pillars of $n = 3.3$. Red line (dash) is the Maxwell Garnett effective medium approximation for the inverted pore medium with the same refractive index contrast. Diamond corresponds to the birefringence from a 2D inverted photonic crystal measured at an optical wavelength of 7 μm [2]. Inset: relative birefringence $\Delta n/n_\perp$ versus volume fraction.

**Figure 4.** (a) Cross-sectional SEM image showing epitaxial wires grown on GaP (100), with (inset) schematic overview of crystallographic growth direction of the nanowires on the GaP substrate. Scale bar represents 1 μm. (b) Scheme of the experimental configuration in which the polarized transmission is measured at normal incidence as a function of the azimuthal rotation of the nanowire material. (c)

Transmitted intensity ratio $I_\perp/I_{//}$ as a function of the azimuthal orientation $\phi$ of the sample, for light impinging perpendicularly to the nanowire composite medium, grown on a GaP (100) substrate.

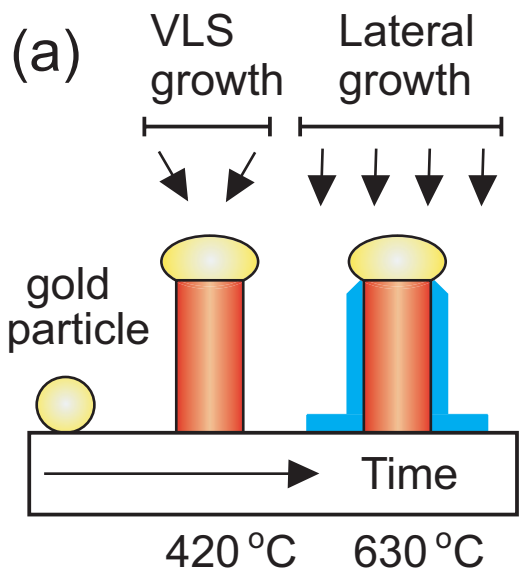
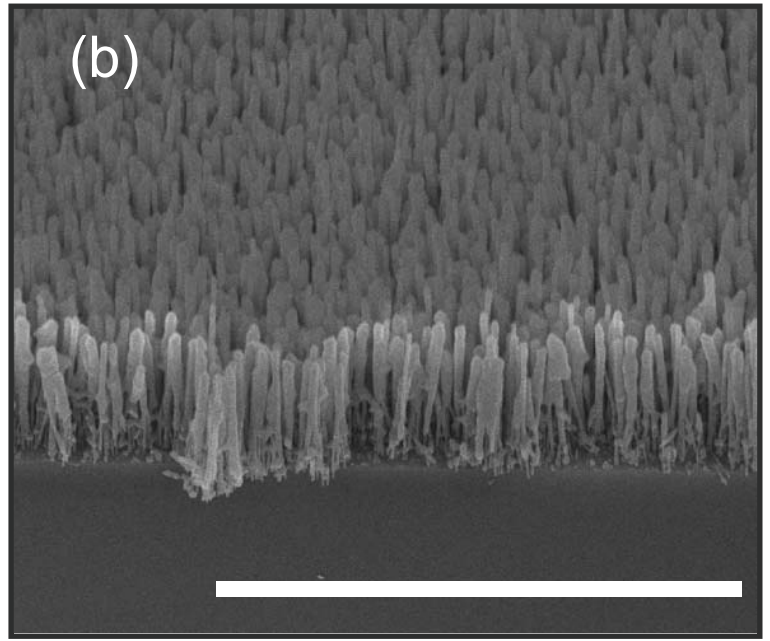
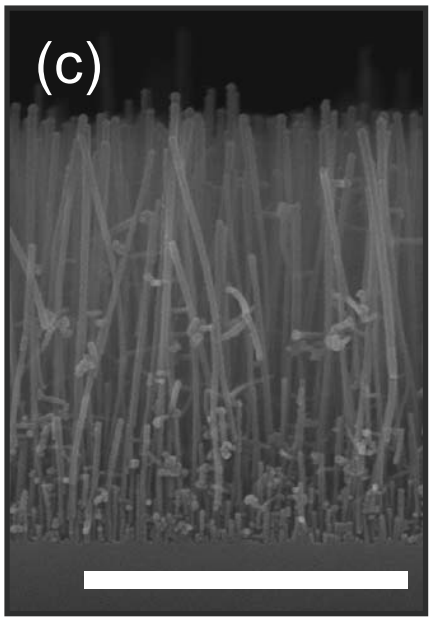
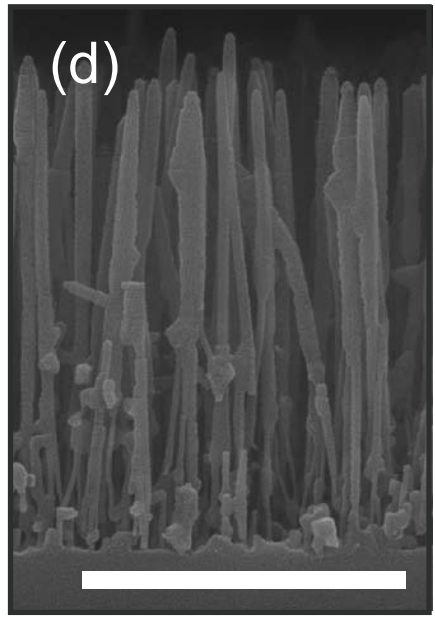
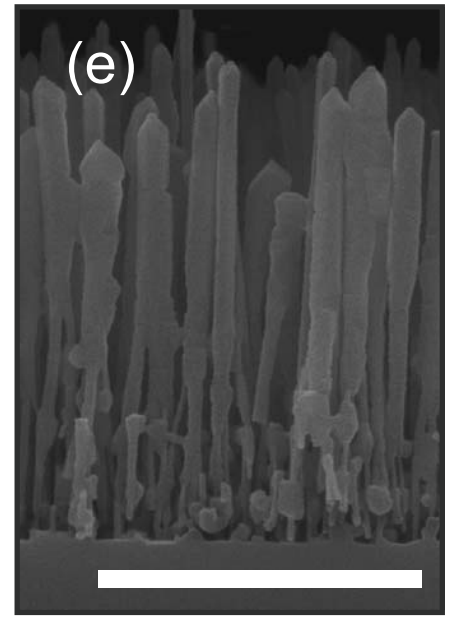

Figure 1.

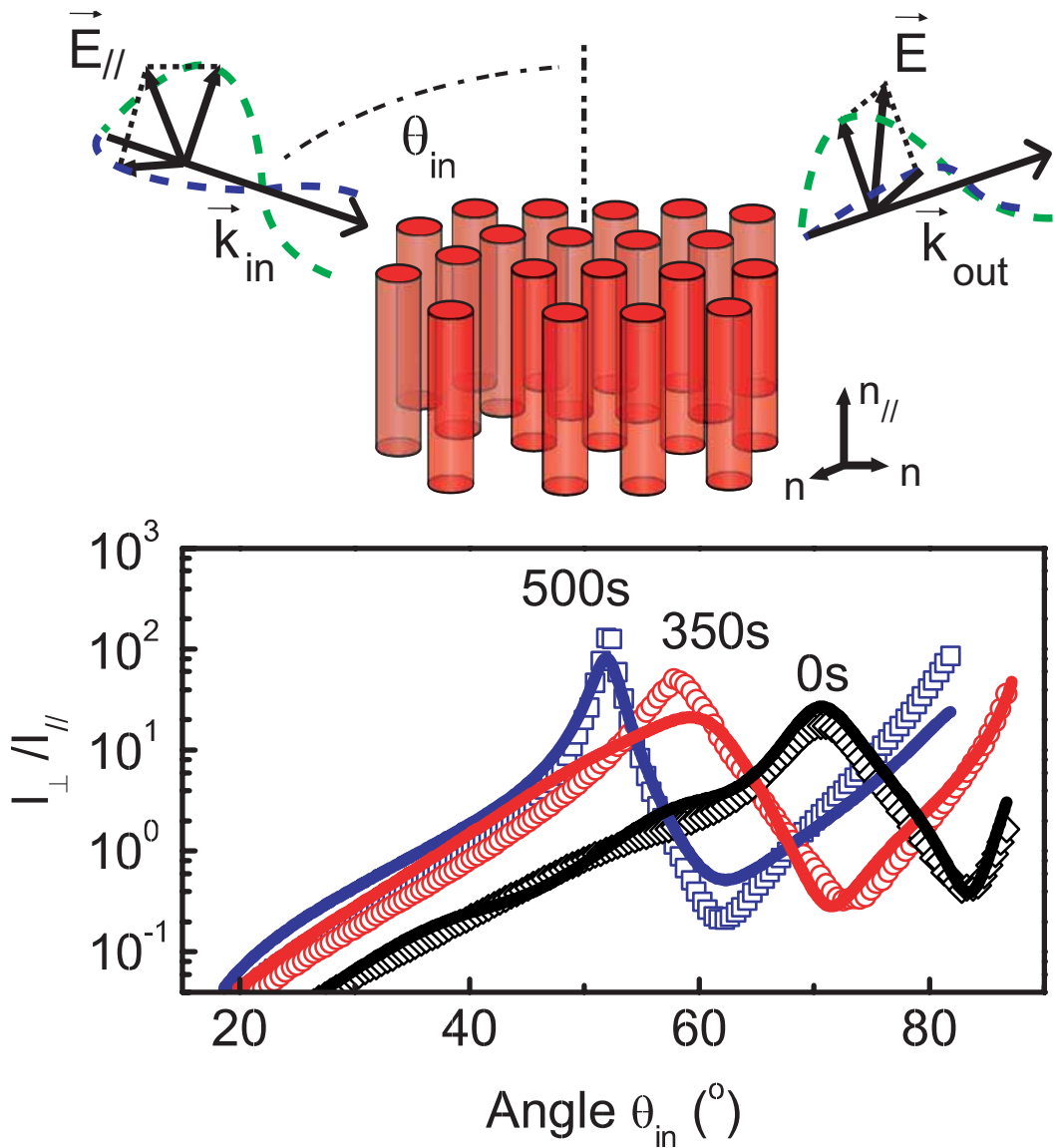

Figure 2.

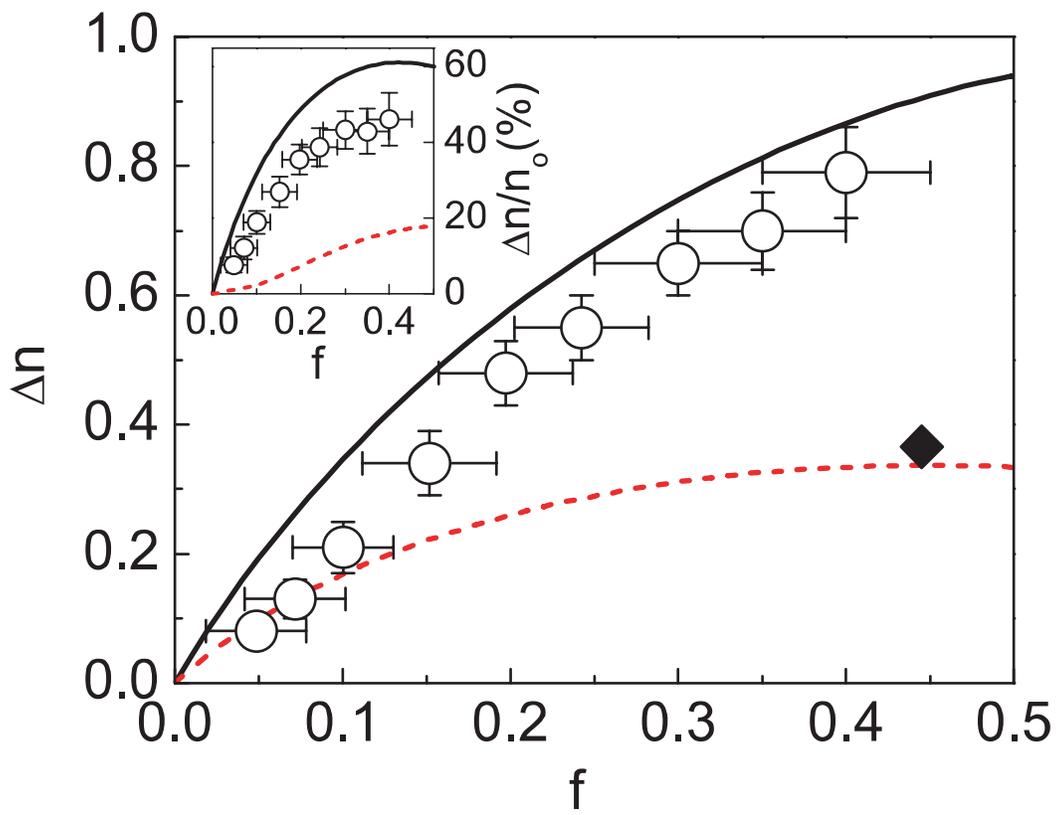

Figure 3.

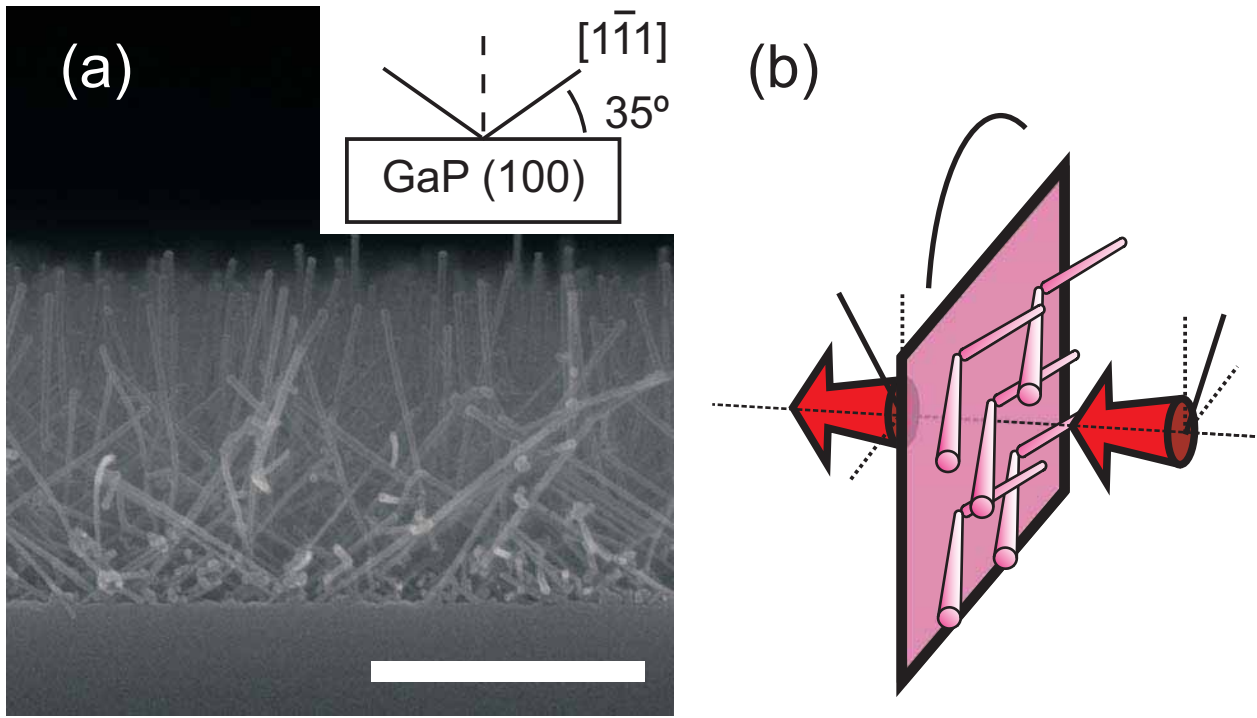
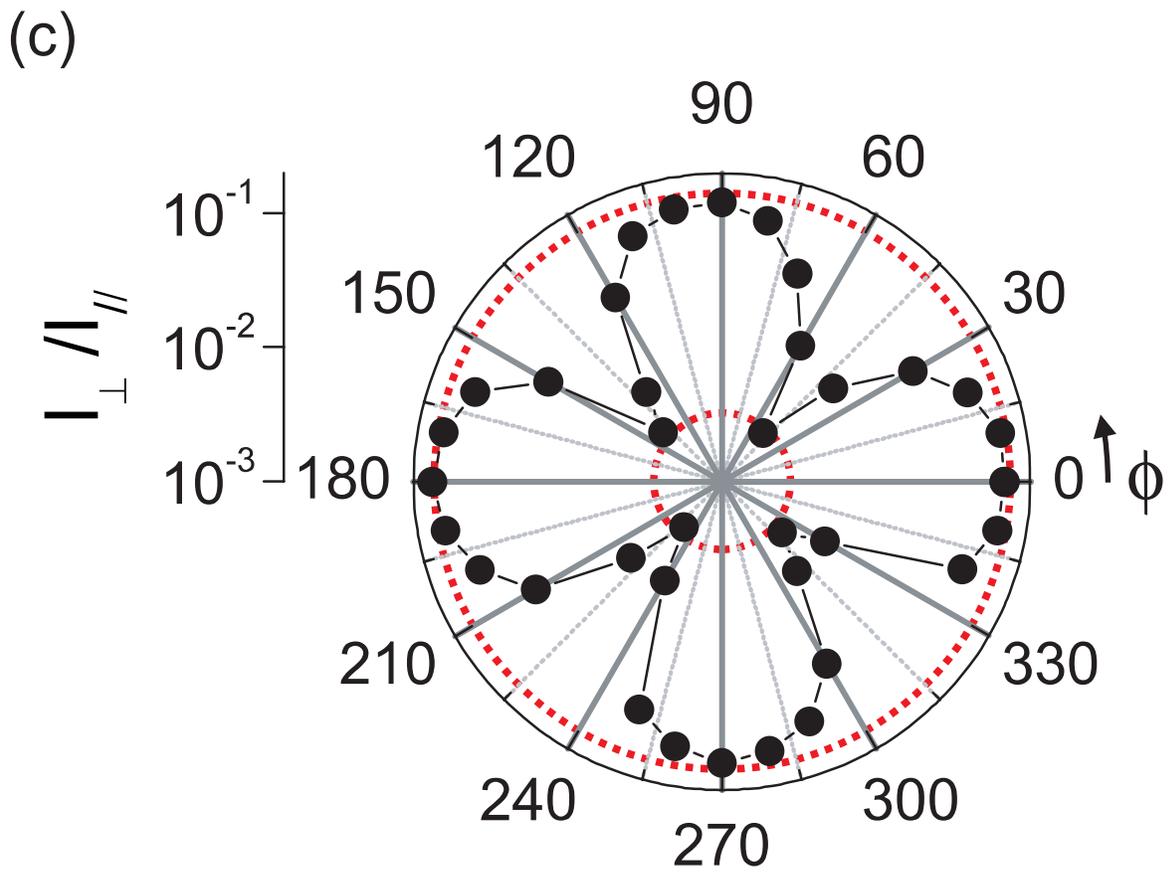

Figure 4.